\documentclass[journal=jctcce,manuscript=article]{achemso}
\setkeys{acs}{chaptertitle=true}
\usepackage[T1]{fontenc}
\usepackage{amsmath, amssymb}
\usepackage{siunitx, chemformula}
\usepackage{multirow}
\usepackage{array}
\usepackage{enumitem}
\usepackage[colorlinks=true, citecolor=blue, urlcolor=blue, linkcolor=blue, breaklinks=true, pdfpagelabels=false]{hyperref}
\usepackage{booktabs}
\usepackage{pdfpages}

\newcommand*\m[1]{\mathrm{#1}}

\newcommand*\annotation[1]{\textsuperscript{\emph{#1}}}
\usepackage{physics}

\author{Wenxian Qin}
\altaffiliation{Contributed equally to this work}
\author{Tai Wang}
\altaffiliation{Contributed equally to this work}
\author{Yue Yu}
\author{Yuanyang Liu}
\author{Yi Qin Gao}
\author{Yunlong Xiao}
\affiliation{College of Chemistry and Molecular Engineering, Peking University, Beijing 100871, The People’s Republic of China}
\email{xiaoyl@pku.edu.cn}

\title{Spin-Consistency Constraints in Noncollinear Tensor TDA}

\begin{document}

\begin{tocentry}
\includegraphics[width=3.2in]{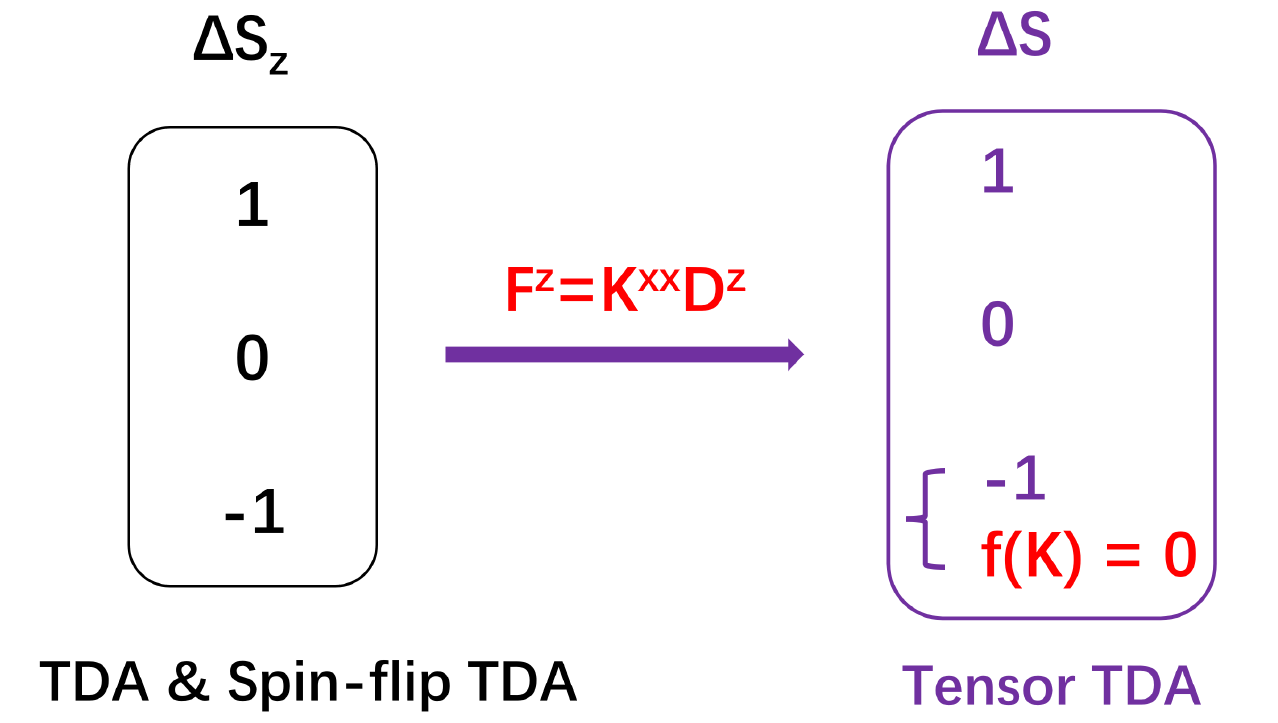}
\end{tocentry}

\begin{abstract}
TDDFT for open-shell systems, whether spin-conserving or spin-flip, has long suffered from spin contamination. This problem arises because the single-excitation space built upon a single Kohn–Sham determinant is not spin-complete. Adopting spin tensor reference states therefore offers an elegant and promising route to resolving this issue. In this work, we revisit the tensor TDDFT equations within the Tamm–Dancoff approximation (TDA) from a noncollinear perspective. We show that, for S = 1/2 reference states, the internal consistency of the spin tensor formulation can, with the aid of the zero-excitation-energy theorem, be recast as a set of constraints that the exchange-correlation kernel must satisfy. Standard noncollinear functionals, however, generally fail to meet these constraints. To address this, we propose a kernel reconstruction scheme that is independent of the specific functional form and free of empirical parameters. This scheme enforces the required constraints, restoring internal consistency in the full spin tensor structure, with spin adaptation following as a natural consequence. Furthermore, when extended to tensor reference states with other values of S, such as S = 1 for the oxygen molecule, the scheme eliminates the so-called artifact states, namely solutions with severely underestimated excitation energies. In addition, the scheme allows the target states that ROKS reference states aim to describe to be expressed and computed, at the TDA level, within the same unified framework as other states, a capability that spin-adapted spin-conserving TDDFT has so far lacked.
\end{abstract}

\maketitle

\section{\label{sec:introduction}Introduction}

Density functional theory (DFT)\cite{hohenberg1964inhomogeneous,kohn1965selfconsistent} is the workhorse electronic-structure method for molecular and materials simulations. Linear-response time-dependent density functional theory (TDDFT)\cite{runge1984densityfunctional,casida1995timedependent} extends this framework to electronic excited-state properties. When spin-orbit coupling (SOC) is neglected, DFT calculations often adopt unrestricted or restricted open-shell Kohn--Sham (UKS or ROKS) formalisms, which are collinear in the sense that a single, fixed spin quantization axis is chosen. Without loss of generality, it is often chosen to be the $z$ axis. For a collinear DFT reference, the TDDFT excitation space separates naturally according to the change in spin projection, $\Delta S_z$.\cite{ullrich2012systems,shao2003spin} We refer to excitations with $\Delta S_z=0$ as spin-conserving excitations, and to those with $\Delta S_z=+1$ and $\Delta S_z=-1$ as spin-flip-up and spin-flip-down excitations, respectively.

For a closed-shell reference, which is the most common choice in practice, the spin-conserving channel is usually sufficient to describe both singlet and triplet excited states without spin contamination\cite{dreuw2005singlereference}. Correspondingly, the spin-flip-up and spin-flip-down excitations give the $S_z=+1$ and $S_z=-1$ components of triplet states, respectively, which should be degenerate with the $S_z=0$ triplet states obtained from the spin-conserving channel\cite{li2023noncollinear,li2025analytic}. We emphasize that throughout this work, we consider only spin-free Hamiltonians that are invariant under global spin rotations and do not include spin-dependent interactions such as external magnetic fields or spin--orbit coupling.

Open-shell references are nevertheless essential in many chemically important situations. For instance, odd-electron systems, including organic and inorganic radicals, are intrinsically open-shell.\cite{li2016critical,li2016criticala} Diradicals\cite{stuyver2019diradicals} and bond-dissociation problems\cite{golubeva2007performance} often involve nearly degenerate frontier orbitals, for which a closed-shell reference can become unstable. In such cases, a high-spin open-shell reference often provides a more robust starting point. Open-shell references may also be chosen intentionally, for example, for thermally activated delayed fluorescence and inverted singlet-triplet gap systems,\cite{desilva2019inverted} for states with double-excitation character relative to a closed-shell reference,\cite{rinkevicius2010spinflipb} or for the description of conical intersections.\cite{minezawa2009optimizing} However, extending TDDFT from closed-shell to open-shell references is not straightforward.

A central difficulty is that the commonly used single-excitation configuration space is not spin-complete. As a result, a naive open-shell TDDFT calculation may produce states that are not eigenfunctions of $\hat{S}^2$ and may suffer from severe spin contamination. At the wave-function level, approaches such as spin-complete spin-flip configuration interaction singles (SC-SF-CIS) \cite{sears2003spincomplete} and   (spin-flip) extended CIS (XCIS and SF-XCIS)\cite{maurice1996nature,casanova2008spinflip}  
address this problem by enlarging the excitation space to include all relevant configurations. Within the TDDFT framework, by contrast, an alternative route is provided by the tensor equation-of-motion (TEOM) formalism.\cite{rowe1975tensor,li2010spinadapted,zhang2015spinflip,chibueze2025spinadapted} In this approach, the three $S_z$-resolved channels are combined to construct Hamiltonian blocks for target states with definite total spin. For references with spin $S\geqslant 1$, these target manifolds are $S_\m{t}=S+1$, $S_\m{t}=S$, and $S_\m{t}=S-1$. Crucially, this construction achieves spin completeness with little or no increase in the size of the configuration space.

However, when directly employing the unmodified TEOM approach, the excitation energy expression is not unique for an $S=1/2$ reference state. This issue has been previously noted by Casida\cite{casida2005propagator} as well as by Li and Liu.\cite{li2011spinadapted} In this work, we provide a systematic analysis within the Tamm--Dancoff approximation (TDA) and show that the origin of this non-uniqueness lies in the fact that the Casida $A$ matrices constructed from different $S_z$ spin channels do not satisfy the required spin-consistency constraints. Importantly, this set of spin-consistency constraints can be reduced, using the zero-excitation-energy theorem proposed in our previous work,\cite{wang2025zero} to conditions involving only the noncollinear exchange-correlation kernel (including both spin-conserving and spin-flip components). By reconstructing the kernel to enforce these constraints exactly, a unique excitation energy expression is recovered for the $S=1/2$ case. We refer to the resulting TDA-level working formulation as noncollinear tensor TDA (NT-TDA), whereas the formulation prior to kernel reconstruction is referred to as the unreconstructed tensor TDA.

Furthermore, for triplet reference states ($S=1$), the problem of anomalously low singlet excitation energies (the ``artifacts'' identified by Chibueze and Visscher\cite{chibueze2025spinadapted}) is also systematically resolved within the same framework. Within this unified framework, spin adaptation for all target states ($S_\m{t} = S-1, S, S+1$) emerges naturally, rather than being treated separately within individual $S_z$ channels. This represents a clear distinction from previous approaches, including X-TDDFT\cite{li2011spinadapted}, XSF-TDA\cite{zhao2026spinadapted}, Q-SF-TDA\cite{chibueze2025spinadapted} and SA-SF-DFT\cite{zhang2015spinflip}. It is worth noting that Zhang and Wang have recently proposed an alternative spin-adapted TDDFT scheme, in which the pure functional contribution is restricted to a spin-unpolarized form.\cite{zhang2026spinadapted}

The remainder of this paper is organized as follows. Section~\ref{sec:theory} presents the theoretical formulation. Section~\ref{sec:applications} presents the numerical results and discussion. Section~\ref{sec:conclusion} presents the conclusions.

\section{\label{sec:theory}Theory}
\subsection{\label{subsec:col_tda}Noncollinear TDA Based on a Single ROKS Reference}

In a two-component spinor basis, the Casida equation within the Tamm–Dancoff approximation (TDA) is given by
\begin{equation}
    A_{AI,BJ}X_{BJ}=\Omega X_{AI},
    \label{eq:casida_tda}
\end{equation}
where we use $I,J,\ldots$ to denote occupied spin orbitals, $A,B,\ldots$ to denote unoccupied spin orbitals, and Einstein summation is implied throughout this work. The ``effective'' Hamiltonian matrix $A$ is defined as
\begin{equation}
    A_{PQ,RS} = F_{PR}\delta_{QS}-\delta_{PR}F_{SQ}+K_{PQ,RS},
    \label{eq:general_A_spinorbital}
\end{equation}
where the one-particle density matrix in the spinor basis is
\begin{equation}
    D_{PQ}=\expval*{a_Q^\dagger a_P}.
    \label{eq:density_matrix_def}
\end{equation}
The Fock ($F$) and kernel ($K$) matrices are defined as the first and second derivatives of the energy $E$ with respect to $D$, respectively:
\begin{equation}
    F_{PQ}=\frac{\partial E}{\partial D_{QP}},\qquad
    K_{PQ,RS}=\frac{\partial^2 E}{\partial D_{QP}\partial D_{RS}}.
    \label{eq:F_K_def}
\end{equation}

To avoid spin contamination in the TDA calculations, we first require a spin-pure reference state. Since noncollinear references and unrestricted Kohn--Sham (UKS) references generally exhibit spin contamination, they are beyond the scope of the present work. Instead, we adopt a high-spin, $S_z$-adapted restricted open-shell Kohn--Sham (ROKS) determinant throughout this work. Each spin orbital in this determinant can therefore be written as a product of a real-valued spatial orbital and a spin function, $P\equiv p\sigma$ with $\sigma=\uparrow,\downarrow$. Accordingly, the two-component quantities introduced above can be resolved into spatial and spin indices,
\begin{equation}
    D^{\sigma\sigma'}_{pq}=D_{p\sigma q\sigma'},\qquad
    F^{\sigma\sigma'}_{pq}=F_{p\sigma q\sigma'},\qquad
    K^{\sigma\sigma',\tau\tau'}_{pq,rs}
    =K_{p\sigma q\sigma',r\tau s\tau'}.
\end{equation}
For such a collinear reference, the spin density matrices are
\begin{equation}
    D^{\uparrow\uparrow}_{pq}=\expval*{a_{q\uparrow}^{\dagger}a_{p\uparrow}},
    \qquad D^{\uparrow\downarrow}_{pq}=D^{\downarrow\uparrow}_{pq}=0, \qquad
    D^{\downarrow\downarrow}_{pq}=\expval*{a_{q\downarrow}^{\dagger}a_{p\downarrow}},
\end{equation}
or, equivalently, in the so-called $0,x,y,z$ representation,
\begin{equation}
D^0=D^{\uparrow\uparrow}+D^{\downarrow\downarrow},\qquad
    D^x=D^{\uparrow\downarrow}+D^{\downarrow\uparrow},
    \qquad
    D^y=i(D^{\uparrow\downarrow}-D^{\downarrow\uparrow}),
    \qquad
    D^z=D^{\uparrow\uparrow}-D^{\downarrow\downarrow}.
\end{equation}
The corresponding Fock matrices are connected via
\begin{equation}
    F^0_{pq}=\pdv{E}{D^0_{qp}}
    =\frac{1}{2}\left(F^{\uparrow\uparrow}_{pq}+F^{\downarrow\downarrow}_{pq}\right), \qquad
    F^z_{pq}=\pdv{E}{D^z_{qp}}
    =\frac{1}{2}\left(F^{\uparrow\uparrow}_{pq}-F^{\downarrow\downarrow}_{pq}\right).
\end{equation}

 The collinear structure also simplifies the kernel matrix by separating it into spin-conserving, spin-flip-up, and spin-flip-down components. The non-vanishing spin-conserving components are $K^{\uparrow\uparrow,\uparrow\uparrow}$, $K^{\downarrow\downarrow,\downarrow\downarrow}$, $K^{\uparrow\uparrow,\downarrow\downarrow}$, and $K^{\downarrow\downarrow,\uparrow\uparrow}$, while the spin-flip-up and spin-flip-down components are $K^{\uparrow\downarrow,\uparrow\downarrow}$ and $K^{\downarrow\uparrow,\downarrow\uparrow}$, respectively. Since real spatial orbitals are used, these two spin-flip kernels are equivalent, $K^{\uparrow\downarrow,\uparrow\downarrow}=K^{\downarrow\uparrow,\downarrow\uparrow}\equiv K^{\mathrm{SF}}$. Given the block-diagonal structure of the Fock and kernel matrices, Eq.~\eqref{eq:general_A_spinorbital} can be expressed as the direct sum of three $S_z$-adapted TDA matrices: the spin-conserving block $A^{\Delta S_z=0}$, the spin-flip-up block $A^{\Delta S_z=+1}$, and the spin-flip-down block $A^{\Delta S_z=-1}$. For a detailed discussion of the block structure of the kernel matrices and the Casida equation, we refer to Ref.~\citenum{li2025analytic}; although that work is formulated for a UKS reference, most of its formalism and expressions carry over directly to the ROKS reference adopted here.

We adopt the following notational conventions throughout this work. We use $p,q,r,s$ for arbitrary spatial orbitals, $i,j,k,l\in \m{C}$ for doubly occupied closed-shell orbitals, $t,u,v,w\in \m{O}$ for singly occupied open-shell orbitals, and $a,b,c,d\in \m{V}$ for virtual orbitals. The open-shell orbitals are conventionally occupied by spin-up ($\uparrow$) electrons, so that a reference with $N_\m{O}$ open-shell orbitals is denoted by $\ket*{S,S_z=S}$ with $S=N_\m{O}/2$. A key relation used in the present work is the exact identity connecting the spin-polarized Fock matrix to the spin-flip kernel,
\begin{equation}
    2F^z_{pq}=\sum_{u\in \m{O}}K^{\m{SF}}_{pq,uu}.
    \label{eq:ZEET}
\end{equation}
Eq.~\eqref{eq:ZEET} follows from the zero-excitation-energy theorem (ZEET) proposed in Ref.~\citenum{wang2025zero}. Eq.~\eqref{eq:ZEET} is specific to the ROKS reference, while a more general expression valid for arbitrary references is given in Ref.~\citenum{wang2025zero}.

\subsection{\label{subsec:spin_adapted_tda} Unreconstructed Tensor TDA Formulation}

In this subsection, we introduce the unreconstructed tensor TDA formulation in a brief and illustrative way. For the full details of the formalism, please refer to Refs.~\citenum{li2010spinadapted, li2011spinadapteda}. Tensor TDA is formulated not only with a spin tensor reference but also in terms of irreducible spin-tensor excitation operators. Specifically, the single-excitation operators are recombined into irreducible spin-tensor operators. The rank-zero operator is
\begin{equation}
    S_{pq}^\dagger = \frac{1}{\sqrt{2}}\left(a_{p\uparrow}^\dagger a_{q\uparrow} + a_{p\downarrow}^\dagger a_{q\downarrow}\right),
\end{equation}
and the three components of the rank-one operator are $T_{pq}^\dagger(m)$, with $m=+1, 0, -1$,
\begin{subequations}
    \begin{align}
                T_{pq}^\dagger(+1)&=-a_{p\uparrow}^\dagger a_{q\downarrow},\\
        T_{pq}^\dagger(0)&=\frac{1}{\sqrt{2}}\left(a_{p\uparrow}^\dagger a_{q\uparrow} - a_{p\downarrow}^\dagger a_{q\downarrow}\right),\\
T_{pq}^\dagger(-1)&=a_{p\downarrow}^\dagger a_{q\uparrow}.
    \end{align}
    \label{eq:tensor_ops}
\end{subequations}
The spin projection of the tensor operator determines the corresponding $S_z$-adapted TDA block. The $m=+1$ and $m=-1$ components generate the spin-flip-up and spin-flip-down blocks, respectively. The $m=0$ block is slightly different because both $T_{pq}^{\dagger}(0)$ and $S_{pq}^{\dagger}$ contribute. 
In the following spin-tensor analysis, we focus on the rank-one operators and use the following notation
\begin{equation}
    A^{\Delta S_z=m}_{pqrs}
    =
    \mel{\m{Ref}}{T_{pq}(m)\hat{A}T_{rs}^{\dagger}(m)}{\m{Ref}}.
    \label{eq:att0_matrix_element}
\end{equation}

The rank-one operators $T_{pq}^{\dagger}(m)$ acting on the high-spin reference $\ket*{S,S_z=S}$ generate states with well-defined $S_z$ but, in general, without a definite eigenvalue of $\hat{S}^2$. The resulting set of states is referred to here as the ``uncoupled'' $S_z$-adapted basis. For $S\geqslant 1$, angular momentum coupling allows the rank-one tensor to generate three target-spin manifolds, $S_\m{t}=S-1, S$, and $S+1$, where $S_\m{t}$ denotes the total spin quantum number of the target state, although the detailed orbital occupation pattern can remove particular components. The Hamiltonian matrix $A$ in the uncoupled $S_z$-adapted basis (Eq.\ref{eq:att0_matrix_element}) can be readily obtained from the definition given in Section~\ref{subsec:col_tda}. Tensor TDA obtains the corresponding $S_\m{t}$-adapted matrices by decomposing these uncoupled states using group-theoretically determined  coefficients. The scalar operator $S_{pq}^{\dagger}$ is simpler: it produces states that remain spin adapted with target spin $S_\m{t}=S$. In this work, the scalar C-to-V excitations are denoted as the CV0 block, and they will be retained and diagonalized together with the $S_\m{t}=S$ spin-adapted block.

\begin{figure}[H]
    \centering
    \includegraphics[width=0.45\textwidth]{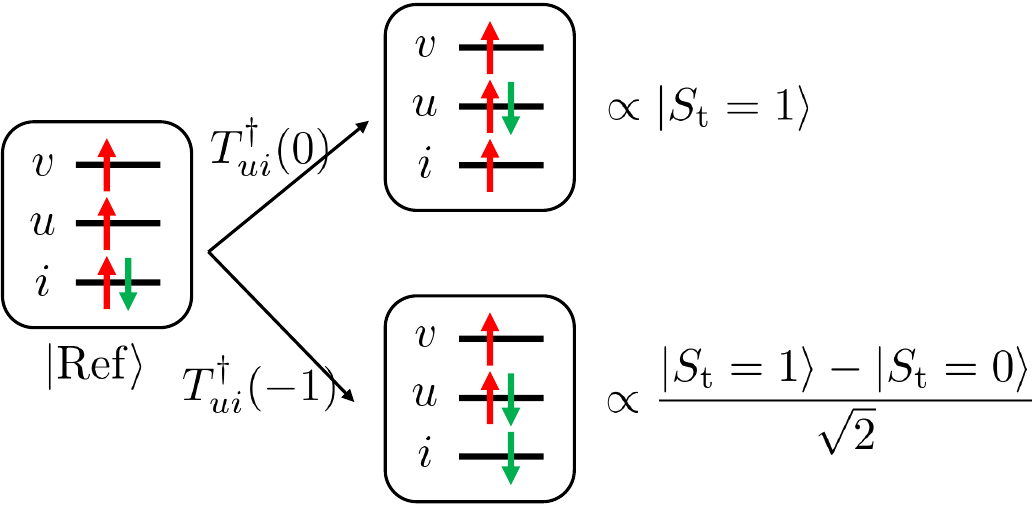}
    \caption{Schematic illustration of the uncoupled $S_z$-adapted states for the C-to-O (CO) excitation on a triplet reference.}
    \label{fig:co_triplet}
\end{figure}

For illustrative purposes, consider a CO excitation on a triplet reference $\ket*{\m{Ref}}\propto \ket*{S=1,S_z=1}$, as illustrated in Fig.~\ref{fig:co_triplet}. The rank-one operators generate the following uncoupled states:
\begin{subequations}
    \begin{align}
        T_{ui}^\dagger(+1) \ket*{\m{Ref}} &= 0,\\
        T_{ui}^\dagger(0) \ket*{\m{Ref}} &= -\frac{1}{\sqrt{2}} \ket*{1,1}, \\
        T_{ui}^\dagger(-1) \ket*{\m{Ref}} &= \frac{1}{\sqrt{2}} \ket*{0,0} - \frac{1}{\sqrt{2}} \ket*{1,0}.
    \end{align}
    \label{eq:ov_decomposition}
\end{subequations}
The states on the left-hand side are expressed in the uncoupled $S_z$-adapted basis, while the same states are represented on the right-hand side as linear combinations of spin-adapted states. Because $\hat{A}$ is a spin scalar, matrix elements between different $S_\m{t}$ sectors vanish. The above decomposition therefore gives, for the diagonal element of the same CO block,
\begin{subequations}
    \begin{align}
        A^{\Delta S_z=0}_{ui,ui} &= \frac{1}{2}A^{S_\m{t}=1}_{ui,ui}, \\
        A^{\Delta S_z=-1}_{ui,ui} &= \frac{1}{2}A^{S_\m{t}=0}_{ui,ui} + \frac{1}{2}A^{S_\m{t}=1}_{ui,ui},
    \end{align}
\end{subequations}
or equivalently,
\begin{subequations}
    \begin{align}
        A^{S_\m{t}=1}_{ui,ui} &= 2A^{\Delta S_z=0}_{ui,ui},\\
        A^{S_\m{t}=0}_{ui,ui} &= 2A^{\Delta S_z=-1}_{ui,ui}-2A^{\Delta S_z=0}_{ui,ui}.
    \end{align}
\end{subequations}
 In this way, the TDA $A$ matrices expressed in the uncoupled $S_z$-adapted basis are transformed into the $S_\m{t}$-adapted basis, and then diagonalized to obtain spin-adapted excitation energies and states. The decompositions of all blocks and the resulting matrix elements used in this work are summarized in Section~S1.1 of the Supporting Information; this construction follows Refs.~\citenum{li2010spinadapted,li2011spinadapteda}.

\begin{figure}[H]
    \centering
    \includegraphics[width=0.45\textwidth]{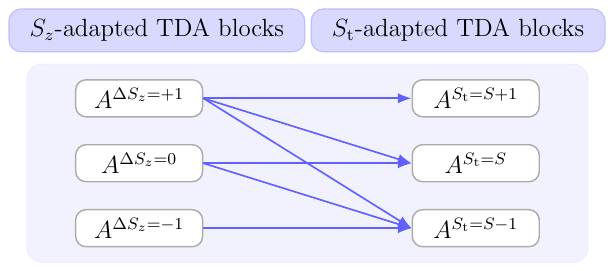}
    \caption{
Transformation between $S_z$-adapted and $S_{\mathrm{t}}$-adapted TDA Hamiltonian blocks.}
    \label{fig:spin_adaptation_cases}
\end{figure}

A key point we wish to emphasize is that the status of this transformation depends critically on the spin of the reference, as summarized in Fig.~\ref{fig:spin_adaptation_cases}.  For $S\geqslant 1$, the three triplet projection channels $\Delta S_z=-1,0,+1$ correspond to the three target-spin sectors $S_\m{t}=S-1,S,S+1$, so the transformation is fixed by Clebsch--Gordan algebra. For a closed-shell reference, $S=0$, the triplet-generated sector contains only $S_\m{t}=1$; the three $S_z$ components are simply the three projections of the same triplet excitation and are trivially degenerate. The doublet case, $S=1/2$, is qualitatively different. The three triplet projection channels can only map onto two target-spin sectors, $S_\m{t}=1/2$ and $S_\m{t}=3/2$. The spin-adapted construction is then overdetermined, and the consistency of the three uncoupled $S_z$-adapted TDA matrices becomes a nontrivial condition. This condition is the focus of the next subsection.

\subsection{\label{subsec:constraints}Spin-Consistency Constraints in NT-TDA}

\begin{figure}[H]
    \centering
    \includegraphics[width=0.45\textwidth]{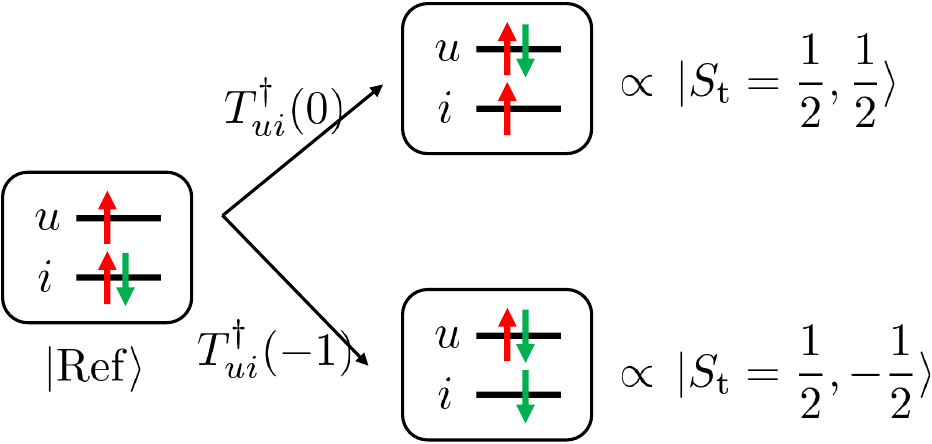}
    \caption{Schematic illustration of the uncoupled $S_z$-adapted states for the CO excitation on a doublet reference.}
    \label{fig:co_doublet}
\end{figure}

We illustrate the doublet constraints with the CO block. As illustrated in Fig.~\ref{fig:co_doublet}, the rank-one tensor operators acting on a doublet reference give
\begin{subequations}
    \begin{align}
        T_{ui}^\dagger(+1)\ket*{\m{Ref}} &= 0,\\
        T_{ui}^\dagger(0)\ket*{\m{Ref}}
        &= -\frac{1}{\sqrt{2}}\ket*{\frac{1}{2},\frac{1}{2}},\\
        T_{ui}^\dagger(-1)\ket*{\m{Ref}}
        &= -\ket*{\frac{1}{2},-\frac{1}{2}}.
    \end{align}
\end{subequations}
The two nonzero states belong to the same $S_\m{t}=1/2$ multiplet. This implies
\begin{equation}
    A^{\Delta S_z=-1}_{ui,ui} = 2A^{\Delta S_z=0}_{ui,ui}.
\end{equation}
It can easily be generalized to
\begin{equation}
    A^{\Delta S_z=-1}_{ui,uj} = 2A^{\Delta S_z=0}_{ui,uj},
    \label{eq:doublet_ov_projection_condition}
\end{equation}
where $i, j$ can be the same or different indices.

The two uncoupled TDA matrix elements can be written directly in terms of the Fock and kernel matrices as
\begin{subequations}
    \begin{align}
        A^{\Delta S_z=0}_{ui,uj}
        &=\frac{1}{2}
        \left( \delta_{ij} F^{\downarrow\downarrow}_{uu}
        -F^{\downarrow\downarrow}_{ji}
        +K^{\downarrow\downarrow,\downarrow\downarrow}_{ui,uj}
        \right),\\
        A^{\Delta S_z=-1}_{ui,uj}
        &=\delta_{ij} F^{\downarrow\downarrow}_{uu}
        -F^{\uparrow\uparrow}_{ji}
        +K^{\downarrow\uparrow,\downarrow\uparrow}_{ui,uj}.
    \end{align}
\end{subequations}
Substitution into Eq.~\eqref{eq:doublet_ov_projection_condition} gives
\begin{equation}
    2F_{ji}^z=K^{\downarrow\uparrow,\downarrow\uparrow}_{ui,uj}-K^{\downarrow\downarrow,\downarrow\downarrow}_{ui,uj}.
\end{equation}
At this point, the relation still involves both the Fock matrix and the kernel. The zero-excitation-energy theorem in Eq.~\eqref{eq:ZEET} allows the spin-dependent part of the Fock matrix,
$F^z$, to be expressed through the spin-flip kernel. The consistency condition can thus be written solely in terms of kernels
\begin{equation}
    K^{\downarrow\downarrow,\downarrow\downarrow}_{ui,uj} = K^{\downarrow\uparrow,\downarrow\uparrow}_{ui,uj} - K^{\downarrow\uparrow,\downarrow\uparrow}_{uu,ij},
    \label{eq:constraint_1}
\end{equation}
for the CO--CO block.

Equation~\eqref{eq:constraint_1} is the CO--CO consistency condition. Among the 16 blocks of the $A$ matrix generated by the four excitation classes CO, CV, OO, and OV, only 10 are independent after hermiticity is taken into account. The OO excitation space is already spin complete; therefore, after applying the zero excitation energy theorem, the consistency conditions involving the OO block are satisfied automatically. The remaining five independent kernel constraints give
\begin{align}
    \m{OV\text{--}OV:}\quad
    K^{\uparrow\uparrow,\uparrow\uparrow}_{au,bu}
    &= K^{\downarrow\uparrow,\downarrow\uparrow}_{au,bu} - K^{\downarrow\uparrow,\downarrow\uparrow}_{ab,uu},
    \label{eq:constraint_2}\\
    \m{OV\text{--}CO:}\quad
    K^{\uparrow\uparrow,\downarrow\downarrow}_{au,uj}
    &=-K^{\downarrow\uparrow,\downarrow\uparrow}_{au,uj},
    \label{eq:constraint_3}\\
    \m{CO\text{--}CV:}\quad
    K^{\downarrow\downarrow,\downarrow\downarrow}_{ui,aj}
    &=K^{\downarrow\downarrow,\uparrow\uparrow}_{ui,aj}
    +K^{\downarrow\uparrow,\downarrow\uparrow}_{ui,aj},
    \label{eq:constraint_4}\\
    \m{OV\text{--}CV:}\quad
    K^{\uparrow\uparrow,\uparrow\uparrow}_{au,bj}
    &=K^{\uparrow\uparrow,\downarrow\downarrow}_{au,bj}
    +K^{\downarrow\uparrow,\downarrow\uparrow}_{au,bj},
    \label{eq:constraint_5}\\
    \m{CV\text{--}CV:}\quad
    K^{\downarrow\uparrow,\downarrow\uparrow}_{ai,bj}
    &=K^{\m{T}}_{ai,bj},
    \label{eq:constraint_6}
\end{align}
where $K^{\m{T}}=\frac{1}{2}(K^{\uparrow\uparrow\uparrow\uparrow}+K^{\downarrow\downarrow\downarrow\downarrow}-K^{\uparrow\uparrow\downarrow\downarrow}-K^{\downarrow\downarrow\uparrow\uparrow})$ is defined in Eq.~(S1.4) of the Supporting Information. Eqs.~\eqref{eq:constraint_1}--\eqref{eq:constraint_6} are therefore the complete set of nontrivial kernel constraints that must hold for the three uncoupled $S_z$-adapted TDA matrices to generate consistent $S_\m{t}$-adapted matrices for an $S=1/2$ reference. This is the central obstruction addressed in the present work. Commonly used density functional approximations do not generally satisfy these identities. Consequently, without further modification, the resulting noncollinear tensor TDA does not possess a genuine tensor structure.

\subsection{\label{subsec:scheme}Spin-Consistent Kernel Reconstruction}

In this subsection, we propose a scheme to restore spin-consistency, namely enforcing the constraints in Eqs.~\eqref{eq:constraint_1}--\eqref{eq:constraint_6}.

The first step of the scheme addresses Eq.~\eqref{eq:constraint_6}, which is the ``spin-degeneracy condition'' analyzed by Li and Liu in Refs.~\citenum{li2010spinadapted,li2011spinadapteda}. 
As shown in previous analyses,\cite{bast2009relativistic,chibueze2025spinadapted,wang2025zero} this condition is automatically satisfied for all functionals in the closed-shell limit, where $D^{\uparrow\uparrow}=D^{\downarrow\downarrow}$, or equivalently, $D^z=0$. Motivated by this observation, we therefore evaluate the kernel using  
\begin{equation}
    D^z \to 0.
    \label{eq:scheme1}
\end{equation}
This choice is further justified by the tensor nature of the reference. For a given total spin $S$, any of the 2S+1 microstates with $S_z=-S,-S+1,...,S$ are equivalent representations of the same state, although the $S_z=S$ determinant is typically adopted for computational convenience. Setting $D^z=0$ therefore removes the dependence of the kernel on a particular spin projection while preserving the total density $D^0$. For a detailed discussion of the numerical implications of this choice, see Section~S1.2 of the Supporting Information. In brief, for most systems Eq.~\eqref{eq:scheme1} has only a minor numerical effect; however, for systems close to response instabilities, it is essential for maintaining numerical stability.

With the spin-unpolarized kernel evaluation above, the spin-channel components obey
\begin{subequations}
    \begin{align}
        K^{\uparrow\uparrow,\uparrow\uparrow}
        &=K^{\downarrow\downarrow,\downarrow\downarrow},\\
        K^{\uparrow\uparrow,\downarrow\downarrow}
        &=K^{\downarrow\downarrow,\uparrow\uparrow},\\
             K^{\uparrow\uparrow,\uparrow\uparrow}   
        &=K^{\uparrow\uparrow,\downarrow\downarrow}+K^{\mathrm{SF}}.
    \end{align}
\end{subequations}
 Consequently, the constraints in Eqs.~\eqref{eq:constraint_4}--\eqref{eq:constraint_6} are automatically satisfied. The remaining constraints in Eqs.~\eqref{eq:constraint_1}--\eqref{eq:constraint_3} reduce to
\begin{subequations}
    \begin{align}
        K^{\uparrow\uparrow,\downarrow\downarrow}_{au,bu} &= - K^{\mathrm{SF}}_{ab,uu}, \\
                K^{\uparrow\uparrow,\downarrow\downarrow}_{ui,uj} &= - K^{\mathrm{SF}}_{uu,ij},\\
        K^{\uparrow\uparrow,\downarrow\downarrow}_{au,uj} &= - K^{\mathrm{SF}}_{au,uj}.
    \end{align}
\end{subequations}
To satisfy these three relations simultaneously, we impose the \emph{index-permutation reconstruction}:
\begin{equation}
    K^{\uparrow\uparrow,\downarrow\downarrow} \to - \hat{P} K^{\mathrm{SF}}, \label{eq:scheme2}
\end{equation}
with the permutation operator $\hat{P}$ defined as $\hat{P}K^{\cdots}_{pq, rs}=K^{\cdots}_{pr, qs}$.

This two-step reconstruction, Eqs.~\eqref{eq:scheme1} and \eqref{eq:scheme2}, expresses all kernel blocks in terms of a single reference kernel
\begin{equation}
    K^{\m{Ref}}
    \equiv
    K^{\mathrm{SF}}(D^z=0)
    =
    K^{\m{T}}(D^z=0),
    \label{eq:Kref_def}
\end{equation}
and is equivalent to the following replacement
\begin{subequations}
    \begin{align}
        K^{\mathrm{SF}}
        &\to K^{\m{Ref}},\\
        K^{\uparrow\uparrow,\downarrow\downarrow}
        &=K^{\downarrow\downarrow,\uparrow\uparrow}
        \to -\hat{P}K^{\m{Ref}},\\
        K^{\uparrow\uparrow,\uparrow\uparrow}
        &=K^{\downarrow\downarrow,\downarrow\downarrow}
        \to (1-\hat{P})K^{\m{Ref}}.
\end{align}        \label{eq:kernel_from_Kref}
\end{subequations}
For a general approximate DFT functional, the reconstructed kernel is not guaranteed to remain the second functional derivative of an exchange-correlation energy functional. The $D_z=0$ evaluation itself does not cause this issue, since the kernel is still obtained by differentiating the underlying functional and evaluating the result at $D_z=0$. The subsequent index-permutation reconstruction, however, is imposed algebraically at the kernel level to enforce the required spin-tensor identities. It is worth noting that this reconstruction does not alter the Hartree--Fock kernel, since the HF kernel intrinsically satisfies the required spin-consistency relations. In this sense, the reconstruction can be viewed as imposing, or mimicking, the spin-tensor structure inherent to the Hartree--Fock kernel.

The explicit expressions of NT-TDA after kernel reconstruction are given in Section~S1.3 of the Supporting Information. The Fock contribution is handled in the same spirit, using the $F^0$ and $F^z$ notation defined in Sec.~\ref{subsec:col_tda}. The scalar component $F^0$ is retained from the ROKS reference, whereas the spin-polarized component is reconstructed from $K^{\m{Ref}}$ through the zero-excitation-energy theorem (Eq.~\eqref{eq:ZEET}):
\begin{equation}
   F^z_{pq}\to \tilde{F}^z_{pq}=\frac{1}{2}\sum_{u\in O}K^{\m{Ref}}_{pq,uu}.
    \label{eq:Fzt_ZEET}
\end{equation}

With the kernel reconstruction in Eqs.~\eqref{eq:kernel_from_Kref}, all six consistency constraints for $S=1/2$ references are satisfied by construction. Although the overdetermination argument is specific to the doublet case, we apply the same kernel reconstruction to all open-shell references. For $S\geqslant 1$, the conventional spin-tensor transformation is formally determined, but practical calculations can still produce low-lying spurious roots for specific $S_\m{t}=S-1$ states, which have been identified as artifacts by Chibueze and Visscher\cite{chibueze2025spinadapted}. Numerical tests in Section~S1.4 of the Supporting Information show that the present reconstruction removes these roots, providing a practical motivation for extending the same kernel construction beyond the doublet case.

Several comments are in order. 
\begin{enumerate}

\item After the above kernel reconstruction is applied, the kernels in different spin channels are no longer treated as independent quantities. Instead, they are all generated from the single reference kernel $K^{\m{Ref}}$ defined in Eq.~\eqref{eq:Kref_def}. The consistency among the $\Delta S_z=0,\pm1$ channels is therefore enforced by construction.

\item The role of the ROKS reference can be understood in terms of the Brillouin condition. Table~S1.7 in the Supporting Information lists the working expressions for the $S_\m{t}=S$ matrix elements after the kernel reconstruction. Inspection of these expressions shows that the reference configuration (OO) has zero diagonal excitation energy in the TDA matrix. However, the ROKS reference does not satisfy all Brillouin conditions\cite{hirao1973general}: in particular, CV-type excitations can still couple to it through nonzero off-diagonal matrix elements. After diagonalization, this coupling gives the reference-dominated $S_\m{t}=S$ state a small downward energy shift relative to the bare ROKS reference. The sign and magnitude of this correction are physically reasonable.

\item Compared with the tensor EOM framework,\cite{rowe1975tensor,li2010spinadapted} the present working equation is closer in spirit to tensor CIS.\cite{zhao2026spinadapted} It avoids the non-Hermitian de-excitation sector present in the EOM formulation, resulting in a simpler eigenvalue problem. This simplification also defines the current scope of the method: the present construction is not straightforwardly extensible to full TDDFT with de-excitation.

\item Finally, the scheme has the correct Hartree--Fock limit and therefore inherits the mild size-consistency error already known for tensor CIS.

\end{enumerate}

\section{\label{sec:applications}Applications}

NT-TDA, together with noncollinear spin-flip TDA and TDDFT, has been implemented
in the NEST (Noncollinear Electronic Structure Theory) software package\cite{nest}, an open-source framework that builds upon the infrastructure provided by PySCF\cite{sun2018pyscf,sun2026python} for developing noncollinear electronic structure methods. All corresponding calculations reported in this work were performed using NEST version 0.1.0.

\subsection{\label{subsec:C2H4}Ethylene torsion}

Ethylene torsion is often used to test electronic-structure methods\cite{shao2003spin,teh2019simplest,zhang2025noncollinear,chakraborty2026new} because the $\pi$ and $\pi^\ast$ frontier orbitals become degenerate at a torsional angle of $90^\circ$. In this geometry, the closed-shell determinant is no longer stable for DFT; therefore it is not a suitable reference for TDDFT. The high-spin triplet configuration, by contrast, remains well defined along the torsional coordinate. TDDFT based on this stable open-shell reference state therefore provides a practical route to the low-lying singlet states.

\begin{figure}[H]
    \centering
    \includegraphics[width=\linewidth]{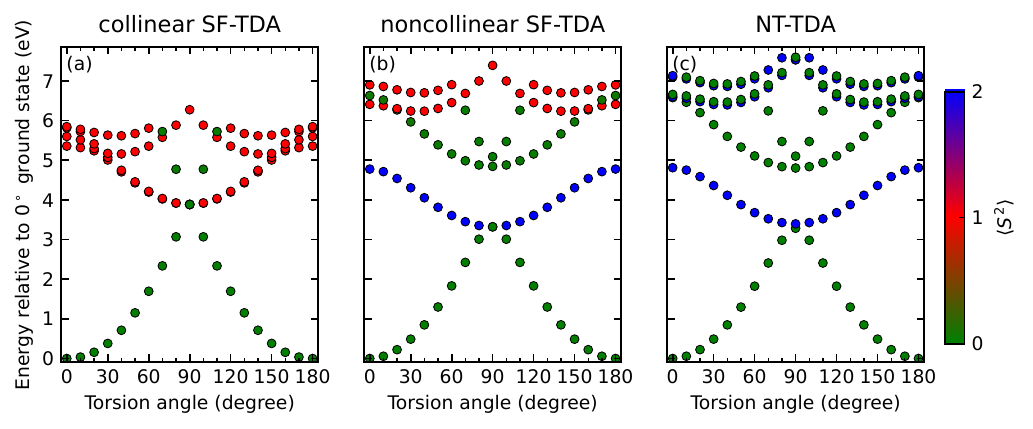}
    \caption{Torsional potential energy curves of ethylene calculated using collinear spin-flip TDA, noncollinear spin-flip TDA, and NT-TDA. The energies are shifted relative to the ground-state energy at a $0^\circ$ torsional angle, and the points are colored by $\expval*{\hat S^2}$. Geometries were taken from Ref.~\citenum{krylov2001sizeconsistent}; calculations were performed at the BLYP/aug-cc-pVTZ level\cite{dunning1989gaussian,kendall1992electron}.}
    \label{fig:C2H4}
\end{figure}

As shown in Fig.~\ref{fig:C2H4}, the collinear spin-flip TDA result obtained with a pure functional such as BLYP\cite{becke1988densityfunctional,lee1988development} is essentially governed by the energy difference between the two frontier orbitals. Consequently, the artificial degeneracy of these orbitals at $90^\circ$ produces a spurious fourfold degeneracy in the torsional potential energy curves. In conventional collinear formulations, hybrid functionals are often employed to alleviate this issue. This artifact can be removed by introducing a noncollinear kernel: noncollinear SF-TDA and NT-TDA give a very similar description at the twisted geometry, and the corresponding vertical excitation energies are listed in Table~S2.1 of the Supporting Information.

However, the advantage of NT-TDA over noncollinear SF-TDA is that it simultaneously eliminates spin contamination. This contamination typically affects only the higher-lying singlets, but in certain cases it can severely affect the $\mathrm{S}_1$ state. The equilibrium geometry of ethylene provides one such example, for which the corresponding vertical excitation energies are listed in Table~S2.2 of the Supporting Information. Eliminating spin contamination is therefore of practical importance, particularly when nonadiabatic coupling between $\mathrm{S}_0$ and $\mathrm{S}_1$ is considered \cite{jing2026analytic}.

\subsection{\label{subsec:Fulvene}Fulvene}

This application is motivated by a recent nonadiabatic dynamics study of ethylene, DMABN, and fulvene using MRSF-TDDFT and MS-CASPT2.\cite{huang2025nonadiabatic} While MRSF-TDDFT showed good agreement with MS-CASPT2 for ethylene and DMABN, the fulvene dynamics exhibited a strong dependence on the choice of density functional. In particular, M06-2X, CAM-B3LYP, and BHHLYP led to substantially different dynamical behaviors. Although analytic gradients and nonadiabatic couplings are not yet available for NT-TDA, we examine here whether NT-TDA provides a more robust static description of fulvene near the relevant nonadiabatic regions. 

We calculate the energies of the two lowest singlet states at four representative geometries: the ground-state minimum $(\m{S}_0)_{\m{min}}$, the first-excited-state minimum $(\m{S}_1)_{\m{min}}$, the minimum-energy conical intersection $(\m{S}_1/\m{S}_0)_{\m{MECI}}$, and a coplanar constrained conical-intersection geometry $(\m{S}_1/\m{S}_0)_{\m{CI}}^{0^\circ}$. These structures were obtained at the XMS-CASPT2/ANO-L level,\cite{segarra-marti2026investigating} and the resulting energies are compared with the corresponding XMS-CASPT2 reference values.
Table~\ref{tab:fulvene} summarizes the results obtained with the def2-SVP basis set\cite{weigend2005balanced} using different functionals\cite{stephens1994initio,becke1993new,yanai2004new,zhao2008m06}. Overall, NT-TDA achieves better agreement with XMS-CASPT2, particularly by reducing the residual energy splitting at the conical intersections for M06-2X, while showing a weaker dependence on the choice of functional. Additional calculations with the def2-TZVP basis set (Table~S2.3 in the Supporting Information) confirm the same trends.

\begin{table}[H]
    \centering
    \caption{Relative energies (eV) as well as the mean absolute error (MAE) of the two lowest singlet states of fulvene at selected geometries calculated with the def2-SVP basis set.}
    \label{tab:fulvene}
    {\small
    \setlength{\tabcolsep}{3pt}
    \begin{tabular}{cc *{8}{>{\centering\arraybackslash}m{1.15cm}}}
    \toprule
    \multirow{2}{*}{Functional} & \multirow{2}{*}{Geometry}              & \multicolumn{3}{c}{MRSF-TDDFT\annotation{a}}                    & \multicolumn{3}{c}{NT-TDA}                                  & \multicolumn{2}{c}{Ref\annotation{b}} \\ \cmidrule(lr){3-5}\cmidrule(lr){6-8}\cmidrule(lr){9-10}
                                &                                        & $\m{S}_0$ & $\m{S}_1$ & $\Delta E_{\m{CI}}$\annotation{c} & $\m{S}_0$ & $\m{S}_1$ & $\Delta E_{\m{CI}}$ & $\m{S}_0$ & $\m{S}_1$ \\ \midrule
    \multirow{5}{*}{B3LYP}      & $(\m{S}_0)_{\m{min}}$                  & 0.00 & 3.25 & --   & 0.00 & 3.22 & --   & 0.00 & 3.25 \\
                                & $(\m{S}_1)_{\m{min}}$                  & 1.03 & 2.73 & --   & 0.96 & 2.45 & --   & 0.90 & 2.38 \\
                                & $(\m{S}_1/\m{S}_0)_{\m{MECI}}$         & 2.45 & 3.17 & 0.73 & 2.51 & 2.58 & 0.07 & 2.38 & 2.38 \\
                                & $(\m{S}_1/\m{S}_0)_{\m{CI}}^{0^\circ}$ & 3.17 & 3.60 & 0.42 & 3.08 & 3.10 & 0.03 & 3.01 & 3.01 \\
                                & MAE\annotation{d}                      & \multicolumn{2}{c}{0.30} & -- & \multicolumn{2}{c}{0.09} & -- & \multicolumn{2}{c}{--} \\ \hline
    \multirow{5}{*}{BHHLYP}     & $(\m{S}_0)_{\m{min}}$                  & 0.00 & 3.33 & --   & 0.00 & 3.34 & --   & 0.00 & 3.25 \\
                                & $(\m{S}_1)_{\m{min}}$                  & 1.32 & 2.49 & --   & 1.08 & 2.44 & --   & 0.90 & 2.38 \\
                                & $(\m{S}_1/\m{S}_0)_{\m{MECI}}$         & 2.52 & 2.57 & 0.05 & 2.45 & 2.69 & 0.24 & 2.38 & 2.38 \\
                                & $(\m{S}_1/\m{S}_0)_{\m{CI}}^{0^\circ}$ & 3.13 & 3.61 & 0.47 & 3.05 & 3.34 & 0.30 & 3.01 & 3.01 \\
                                & MAE\annotation{d}                      & \multicolumn{2}{c}{0.24} & -- & \multicolumn{2}{c}{0.16} & -- & \multicolumn{2}{c}{--} \\ \hline
    \multirow{5}{*}{CAM-B3LYP}  & $(\m{S}_0)_{\m{min}}$                  & 0.00 & 3.27 & --   & 0.00 & 3.27 & --   & 0.00 & 3.25 \\
                                & $(\m{S}_1)_{\m{min}}$                  & 1.18 & 2.66 & --   & 1.04 & 2.45 & --   & 0.90 & 2.38 \\
                                & $(\m{S}_1/\m{S}_0)_{\m{MECI}}$         & 2.46 & 2.95 & 0.49 & 2.53 & 2.60 & 0.07 & 2.38 & 2.38 \\
                                & $(\m{S}_1/\m{S}_0)_{\m{CI}}^{0^\circ}$ & 3.42 & 3.46 & 0.04 & 3.08 & 3.25 & 0.17 & 3.01 & 3.01 \\
                                & MAE\annotation{d}                      & \multicolumn{2}{c}{0.30} & -- & \multicolumn{2}{c}{0.13} & -- & \multicolumn{2}{c}{--} \\ \hline
    \multirow{5}{*}{M06-2X}     & $(\m{S}_0)_{\m{min}}$                  & 0.00 & 3.17 & --   & 0.00 & 3.28 & --   & 0.00 & 3.25 \\
                                & $(\m{S}_1)_{\m{min}}$                  & 1.31 & 2.27 & --   & 1.01 & 2.55 & --   & 0.90 & 2.38 \\
                                & $(\m{S}_1/\m{S}_0)_{\m{MECI}}$         & 2.26 & 2.63 & 0.37 & 2.52 & 2.67 & 0.15 & 2.38 & 2.38 \\
                                & $(\m{S}_1/\m{S}_0)_{\m{CI}}^{0^\circ}$ & 2.86 & 3.57 & 0.71 & 3.17 & 3.21 & 0.04 & 3.01 & 3.01 \\
                                & MAE\annotation{d}                      & \multicolumn{2}{c}{0.24} & -- & \multicolumn{2}{c}{0.16} & -- & \multicolumn{2}{c}{--} \\ \bottomrule
    \end{tabular}}

    \annotation{a}MRSF-TDDFT values were calculated using the OpenQP package\cite{mironov2024openqp,openqp}. \\
    \annotation{b}Reference values are from XMS-CASPT2/ANO-L calculations reported in Ref.~\citenum{segarra-marti2026investigating}. \\
    \annotation{c}$\Delta E_{\m{CI}}=E(\m{S}_1)-E(\m{S}_0)$ at each conical-intersection geometry. \\
    \annotation{d}The MAE is calculated over all listed relative energies except the zero value at $(\m{S}_0)_{\m{min}}$.
\end{table}

We emphasize that these results only assess the static energy description of NT-TDA at the fixed XMS-CASPT2 geometries and do not by themselves demonstrate the ability of NT-TDA to locate conical intersections or describe nonadiabatic dynamics through them.

\subsection{\label{subsec:radicals}Doublet Radicals}

In this benchmark, we investigate doublet radicals using an S = 1/2 reference. The QUEST\#4 benchmark set established by Loos and co-workers\cite{veril2021questdb} is adopted after excluding the six high-symmetry linear molecules CH, CON, NCO, CNO, NO, and OH, because their doublet reference solutions exhibit symmetry breaking and therefore do not preserve the expected orbital degeneracies. The doublet–doublet excitation energies are compared among conventional spin-conserving TDA, noncollinear spin-flip TDA, and NT-TDA. Six density functionals were considered: BLYP, B3LYP, BHHLYP, CAM-B3LYP, PBE0\cite{adamo1999reliable}, and TPSS\cite{tao2003climbing}. Spin-conserving TDA and noncollinear spin-flip TDA use a UKS reference, whereas NT-TDA uses a ROKS reference. For spin-conserving TDA, the reported excitation energies are taken directly from the TDA output. For noncollinear spin-flip TDA and NT-TDA, the lowest TDA root is identified as the ground state, and the excitation energies of the remaining states are evaluated relative to it. The results are summarized in Table~\ref{tab:radicals_MAE}, with the detailed excitation energies provided in Table~S2.4 of the Supporting Information. Overall, the three methods give comparable accuracy for the doublet excitation energies, while only NT-TDA preserves spin consistency.

\begin{table}[H]
    \centering
    \caption{MAE (eV) of vertical excitation energies for doublet radicals calculated using spin-conserving TDA, noncollinear spin-flip TDA, and NT-TDA.\annotation{a}}
    \label{tab:radicals_MAE}
    \begin{tabular}{cccc}
        \toprule
        Functional & spin-conserving TDA & noncollinear spin-flip TDA & NT-TDA \\ \midrule
        BLYP       & 0.47 & 0.62 & 0.60 \\
        B3LYP      & 0.29 & 0.40 & 0.38 \\
        BHHLYP     & 0.28 & 0.32 & 0.18 \\
        CAM-B3LYP  & 0.21 & 0.31 & 0.25 \\
        PBE0       & 0.27 & 0.30 & 0.28 \\
        TPSS       & 0.42 & 0.38 & 0.32 \\ \bottomrule
    \end{tabular}

    \annotation{a}All calculations were performed using the aug-cc-pVTZ basis set. TBE values and geometries were taken from the benchmark set by Loos and co-workers\cite{veril2021questdb}.
\end{table}

\subsection{\label{subsec:NQ}Naphthoquinone Diradicals}

Naphthoquinone diradicals provide an additional test of whether the present scheme can describe adiabatic singlet--triplet ordering in systems with pronounced open-shell character. Photoelectron spectroscopy has confirmed that different naphthoquinone isomers have different spin ground states: 1,8-naphthoquinone has a singlet ground state, whereas 2,7-naphthoquinone has a triplet ground state.\cite{yang2018negative,yang2019negative} We therefore calculated the adiabatic singlet--triplet gap for these two isomers using NT-TDA. For each isomer, the singlet and triplet energies were evaluated separately at the corresponding CASPT2/aug-cc-pVDZ optimized geometries reported previously,\cite{hrovat2018calculations,hrovat2019calculations} and the NT-TDA calculations were performed with the aug-cc-pVTZ basis set using different functionals\cite{dirac1930note,vosko1980accurate,perdew1996generalized}.

As shown in Table~\ref{tab:NQ_ST}, NT-TDA gives the correct qualitative ordering for both isomers with all tested functionals\cite{dirac1930note,vosko1980accurate}: the calculated gaps are negative for 1,8-naphthoquinone and positive for 2,7-naphthoquinone. The magnitude of the gap still depends on the functional, especially for 2,7-naphthoquinone, where pure functionals underestimate the reference gap and hybrid functionals give larger values.

\begin{table}[H]
    \centering
    \caption{Adiabatic singlet--triplet gaps ($\Delta E_{\m{ST}}=E(\m{S})-E(\m{T})$, eV) of naphthoquinone diradicals calculated using NT-TDA.}
    \label{tab:NQ_ST}
    \begin{tabular}{ccc}
        \hline
        Functional & 1,8-naphthoquinone & 2,7-naphthoquinone \\ \hline
        SVWN     & $-0.22$ & 0.27 \\
        BLYP     & $-0.49$ & 0.29 \\
        PBE      & $-0.53$ & 0.29 \\
        B3LYP    & $-0.28$ & 0.45 \\
        PBE0     & $-0.24$ & 0.49 \\
        BHHLYP   & $-0.15$ & 0.61 \\
        Ref\annotation{a} & $-0.13$ & 0.56 \\ \hline
    \end{tabular}

    \annotation{a}Reference values at CASPT2/aug-cc-pVTZ level of theory are reported in Refs.~\citenum{hrovat2018calculations,hrovat2019calculations}.
\end{table}

\subsection{\label{subsec:INVEST}Inverted singlet-triplet gap systems}

In recent years, luminescent materials exhibiting an inverted singlet–triplet energy gap (INVEST) have attracted considerable attention in organic light-emitting diodes (OLEDs). Unlike conventional thermally activated delayed fluorescence (TADF) emitters, INVEST molecules exhibit an inverted excited-state ordering with $\Delta E_{\m{ST}}=E(\m{S}_1)-E(\m{T}_1)<0$, corresponding to the $T_1$ state lying above the $S_1$
 state. This unusual ordering challenges the conventional expectation based on Hund's rule.

\begin{figure}[H]
    \centering
    \includegraphics[width=\linewidth]{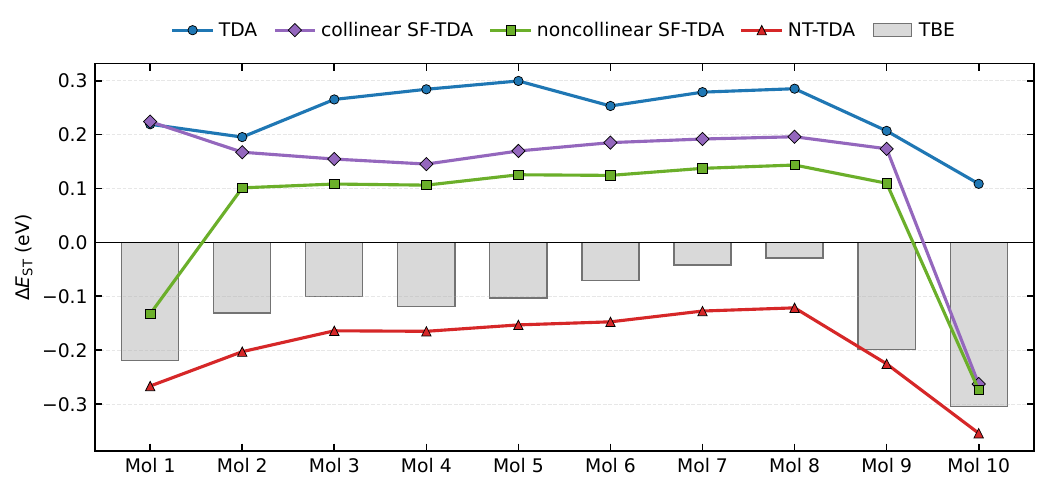} 
    \caption{Comparison of calculated singlet--triplet energy gaps ($\Delta E_{\m{ST}}=E(\m{S}_1) - E(\m{T}_1)$) with TBE values for the ten molecules, calculated at the B3LYP/aug-cc-pVTZ level of theory.}
    \label{fig:STG}
\end{figure}

In this work, we employ the benchmark set of ten triangulene molecules and derivatives established by Loos and co-workers \cite{loos2023heptazine} to assess the performance of spin-conserving TDA, collinear and noncollinear spin-flip TDA, and NT-TDA in predicting singlet–triplet gaps. As shown in Fig.~\ref{fig:STG}  (the numerical data are provided in Table~S2.5 of the Supporting Information), conventional spin-conserving TDA systematically predicts positive singlet--triplet gaps, consistent with previous reports.\cite{desilva2019inverted,lashkaripour2025addressing,zhang2025noncollinear} Spin-flip TDA, whether collinear or noncollinear, predicts inverted gaps for only a few molecules and exhibits strong spin contamination. Among all the tested methods, NT-TDA performs best, predicting negative gaps for all ten molecules. Although Fig.~\ref{fig:STG} shows the B3LYP results, calculations with a broader range of density functionals are provided in Table~S2.6 of the Supporting Information and lead to the same overall conclusions.

\section{\label{sec:conclusion}Conclusion}
We propose a spin-consistent reconstruction scheme of the exchange-correlation kernel within a noncollinear tensor TDDFT framework, which restores internal consistency among different $\Delta S_z$ channels under the TDA approximation. The scheme is applicable to ROKS reference states of arbitrary spin. The spin-consistent reconstruction is formulated independently of the explicit form of the exchange-correlation functional. Compared with conventional collinear spin-flip formulations, NT-TDA also shows reduced sensitivity to the choice of functional. Numerical results demonstrate that the proposed scheme eliminates the low-lying spurious states (the “artifact” identified by Chibueze and Visscher\cite{chibueze2025spinadapted}) present in the unreconstructed formulations, and exhibits robust performance across a range of test systems, including radicals, diradicals, ethylene torsion, INVEST systems, and fulvene.

Unlike conventional spin-adaptation approaches based on spin-conserving and spin-flip excitation pathways, the present scheme: 
\begin{itemize}
    \item treats spin-conserving and spin-flip channels on an equal footing and reconstructs their consistency within a unified framework, thereby yielding a single, spin-consistent set of working equations for the $S_{\mathrm{t}} = S-1$, $S$, and $S+1$ target states;
    \item is free of empirical parameters and independent of the specific functional form, with spin adaptation emerging naturally as a consequence of theoretical self-consistency rather than being imposed explicitly;
    \item allows the target state that a ROKS reference aims to describe to be expressed and computed, at the TDA level, within the same unified framework as all other target states;
    \item achieves removal of spin contamination without requiring explicit projection of the redundant $S_{\mathrm{t}} = S$ spin component out of the OO excitation space for the $S_{\mathrm{t}}=S-1$ calculation, since this component is shown to decouple automatically and appear only as a zero-excitation-energy state within the reconstructed framework. See Section~S1.5 of the Supporting Information for more details.
\end{itemize}

Since the scheme only requires replacing the exchange-correlation kernel in the original working equations, without modifying the tensor TDA framework, it can be readily extended to analytic energy gradients,\cite{furche2002adiabatic,seth2011timedependent,wang2020analytic} nonadiabatic derivative couplings (NADC),\cite{wang2021nactddft,wang2026analytic} and spin-orbit coupling (SOC).\cite{li2013combining,chibueze2026restricted} This unified and spin-consistent framework is particularly advantageous for NADC and SOC calculations, where a consistent treatment of different target-spin manifolds and reduced spin contamination are highly desirable. It is worth noting that the $D_z=0$ prescription can in principle also be applied to the $F^0$ term, although this extension is beyond the scope of the present work.

\begin{acknowledgement}
This work has been supported by Smart Grid-National Science and Technology Major Project (2025ZD0808600 and 2025ZD0808601).
\end{acknowledgement}

\section*{Data Availability}

The NEST software package (version 0.1.0) used to perform the calculations reported in this work is openly available at \url{https://github.com/NonDFT/NEST}. All input and output files required to reproduce the results presented in this manuscript are available in the NEST-publication-data repository at \url{https://github.com/NonDFT/NEST-publication-data/tree/main/NT-TDA}.

\section*{Notes}
The authors declare no competing financial interest.

\begin{suppinfo}
The published Supporting Information is appended after the references and TOC graphic. It is also available at \url{https://doi.org/10.1021/acs.jctc.6c01470}.
\begin{itemize}
    \item \textbf{Part I (Theoretical Details):} Unreconstructed NT-TDA formulation, numerical assessment of the spin-unpolarized kernel reconstruction, reconstructed NT-TDA matrix elements, analysis of the O$_2$ artifact state, and OO-type excitations.
    \item \textbf{Part II (Supplementary Data):} Excitation energies of ethylene; relative energies of fulvene calculated with the def2-TZVP basis set; detailed doublet--doublet excitation energies of doublet radicals; and singlet--triplet gaps of the INVEST molecules calculated using different methods and functionals.
\end{itemize}
\end{suppinfo}
\bibliography{ref}

\AtEndDocument{
    \includepdf[pages=-]{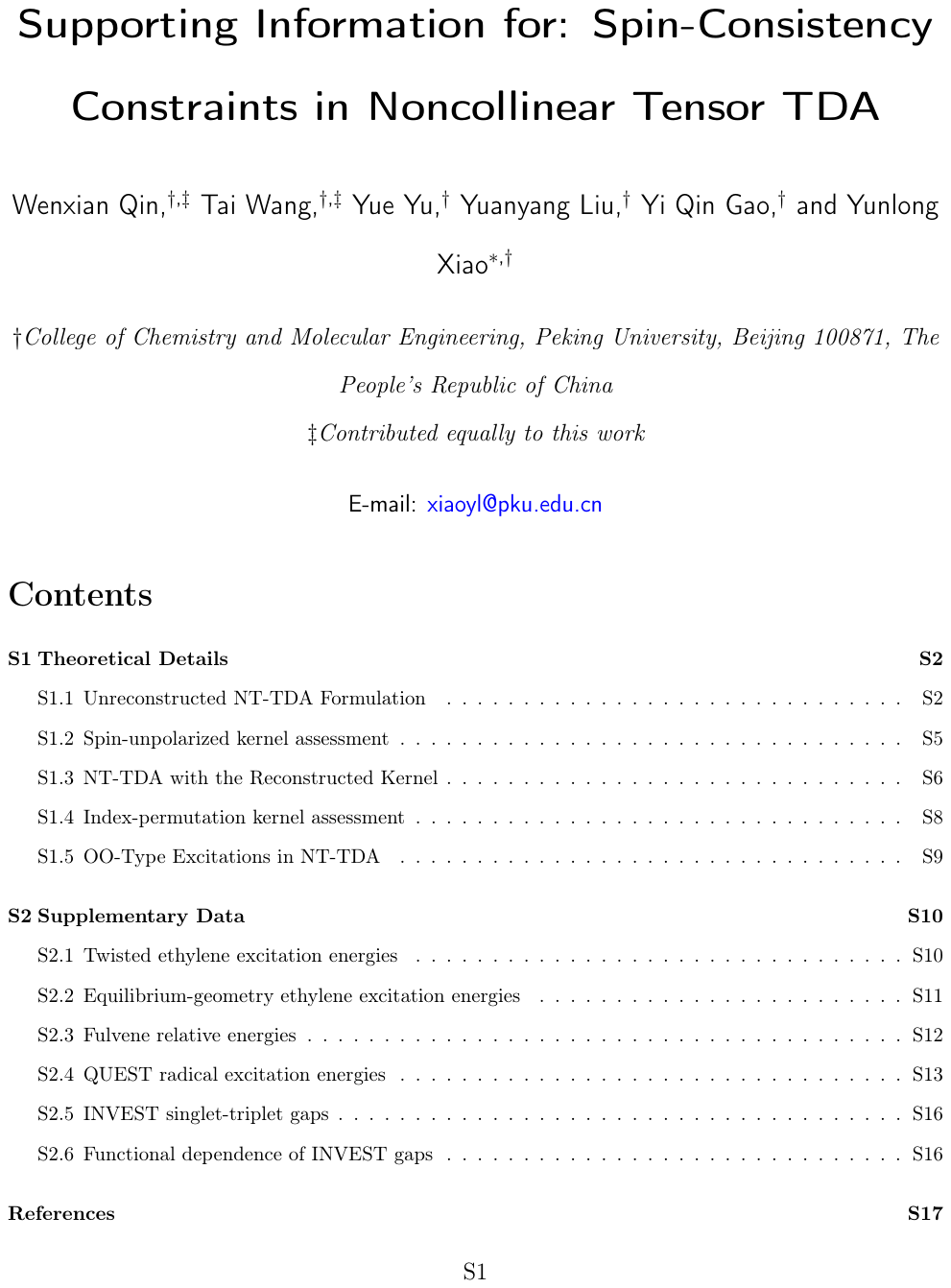}
}

\end{document}